\documentclass[aps,pra,preprint,superscriptaddress,showpacs]{revtex4}
\usepackage{graphicx}
\usepackage{bm}
\begin{document}

\title{Dependence of interface conductivity on relevant physical parameters in polarized Fermi mixtures}

\author{N. Ebrahimian}
\email{n.ebrahimian@aut.ac.ir}
\affiliation{Physics Department, Amirkabir University of Technology, Tehran 15914, Iran}
\author{M. Mehrafarin}
\email{mehrafar@aut.ac.ir}
\affiliation{Physics Department, Amirkabir University of Technology, Tehran 15914, Iran}
\author{R. Afzali}
\email{afzali@kntu.ac.ir}
\affiliation{Physics Department, K.N. Toosi University of Technology, Tehran 15418, Iran}

\begin{abstract}
We consider a mass-asymmetric polarized Fermi system in the presence of Hartree-Fock (HF) potentials.
We concentrate on the BCS regime with various
interaction strengths and numerically obtain the allowed values
of the chemical and HF potentials, as well as the mass ratio. The functional dependence of the heat conductivity of the N-SF
interface on relevant physical parameters, namely the temperature, the mass ratio, and the interaction strength, is obtained. In particular, we show that the interface conductivity starts to drop with decreasing temperature at the temperature, $T_{\text{m}}$, where the mean kinetic energy of the particles is just sufficient to overcome the SF gap. We obtain $T_{\text{m}}$ as a function of the mass ratio and the interaction strength. The variation of the heat conductivity, at fixed temperature, with the HF potentials and the imbalance chemical potential is also obtained. Finally, because the range of relevant temperatures increases for larger values of the mass ratio, we consider the $^6\text{Li}$-$^{40}\text{K}$ mixture separately by taking the temperature dependence of the pair potential into account. 
\end{abstract}

\maketitle

\section{Introduction}
Recently, the study of the behavior of
 ultracold Fermi gases with two imbalanced hyperfine states has opened up an interesting new area in
 many-body atomic physics. In this connection, extensive studies have been
reported that propose various candidates for the pairing state,
including the Fulde-Ferrell-Larkin-Ovchinnikov (FFLO) state
\cite{Fulde,Larkin}, the BCS-normal phase separation
\cite{Bedaque}, the Sarma state \cite{Liu}, the p-wave pairing
state \cite{Petit,Bulgac} and the deformed Fermi surface
superfluid \cite{Muther}. Of central importance is the phase
separation of a superfluid (SF) paired core surrounded by a
polarized normal (N) phase, where, in addition to theoretical
studies
\cite{Silva,Shim,Baur,Caldas,Carlson,Mizushima,Chevy,Haque,Son,Lobo,Sheehy,Ying},
important experimental work has been carried out
\cite{Partridge1,Zwerlein,Shin,Partridge2}. Such a
phase-separation scenario had been proposed by Clogston
\cite{Clogston} and Chandrasekhar \cite{Chandraskhar} long ago,
who predicted the occurrence of a first-order transition from the
N to the SF state. An  interesting result in this connection is
the appearance of a temperature difference between the two phases
as a consequence of the blockage of energy transfer across the
N-SF interface. This blockage is due to a SF gap, which causes
low-energy normal particles to be reflected from the interface. By
studying particle scattering off the interface, the heat
conductivity has been calculated
\cite{Lazaridess,Lazaridess1,Ebrahimian}. 

In this paper we consider a polarized Fermi system consisting of
two spin species with unequal masses in the presence of HF potentials.
We concentrate on the BCS regime with various
interaction strengths and numerically obtain the allowed values
of the chemical and HF potentials, as well as the mass ratio $m_r$. The functional dependence of the heat conductivity of the N-SF
interface on the relevant physical parameters, namely the temperature, the mass ratio, and the interaction strength, is studied in detail. Our focus is on energies slightly above the transmission threshold, because we are
considering low temperatures. In our calculations, we therefore use the approximate low-temperature form of the Fermi-Dirac distribution and regard the pair potential as temperature-independent. In particular, we show that the interface conductivity starts to drop with decreasing temperature at the temperature, $T_{\text{m}}$, where the mean kinetic energy of the particles is just sufficient to overcome the SF gap. The drop is, thus, a result of the blockage of the energy transfer due to the reflection of particles from the interface and signifies a build-up of temperature difference across the interface. We obtain $T_{\text{m}}$ as a function of the mass ratio and the interaction strength. The variation of the heat conductivity, at fixed temperature, with the HF potentials and the imbalance chemical
potential is also obtained. Finally, we single out the particular case of the $^6\text{Li}$-$^{40}\text{K}$
mixture ($m_r=6.7$), due to its importance in experimental and
theoretical studies \cite{Wille,Gubbels,Baarsma}. Because the range of relevant temperatures increases for larger values of the mass ratio, here we take the pair potential to be temperature-dependent and use the exact Fermi-Dirac distribution instead of its approximate low-temperature form. 

\section{Hartree-Fock potential and mass asymmetry effects}
We consider a polarized Fermi gas consisting of two fermionic
species (imbalanced hyperfine states $\uparrow,\downarrow$) of
masses $m_\uparrow, m_\downarrow$ and chemical potentials
$\mu_{\uparrow},\mu_{\downarrow}$ at sufficiently low temperature.
The $\uparrow-\downarrow$ interaction is assumed to be a contact
interaction characterized by the coupling constant $V=-4\pi
a/m_{R}$ ($\hbar=1$) with
$m_{R}=2m_{\uparrow}m_{\downarrow}/(m_{\uparrow}+m_{\downarrow})$.
For the superfluid phase we define the species-imbalance chemical
potential, $h_{s}=(\mu_{\uparrow}-\mu_{\downarrow})/2$, and the average chemical potential,
$\mu_{s}=(\mu_{\uparrow}+\mu_{\downarrow})/2-U_{s}$, where $U_s$
is the superfluid HF potential. For the
calculation of transmission coefficients, and hence the heat
conductivity, we need to obtain the solution of Bogoliubov-de
Gennes equations \cite{deGennes}. The effective hamiltonian of
the system may be written as
\begin{eqnarray}
H=\int
d^3 x\sum_{i}[\psi^{\dagger}(\bm{r}i)H(\bm{r}i)\psi(\bm{r}i)+
U(\bm{r}i)\psi^{\dagger}(\bm{r}i)\psi(\bm{r}i)\nonumber\\
+\Delta(\bm{r})\psi^{\dagger}(\bm{r}\uparrow)\psi^{\dagger}(\bm{r}\downarrow)+\Delta^{\star}(\bm{r})\psi(\bm{r}\downarrow)\psi(\bm{r}\uparrow)]
\end{eqnarray}
where $H(\bm{r}i)=-\frac{\bm{\nabla}^{2}}{2m_{i}}-\mu_{i}$ ($i=\uparrow,\downarrow$) and
\begin{eqnarray}
U(\bm{r}\uparrow)=-V<\psi^{\dagger}(\bm{r}\downarrow)\psi(\bm{r}\downarrow)>
, \ \
U(\bm{r}\downarrow)=-V<\psi^{\dagger}(\bm{r}\uparrow)\psi(\bm{r}\uparrow)>\nonumber\\
\Delta(\bm{r})=-V<\psi(\bm{r}\downarrow)\psi(\bm{r}\uparrow)>=V<\psi(\bm{r}\uparrow)\psi(\bm{r}\downarrow)>\ \ \ \ \ \ \ \ \nonumber
\end{eqnarray}
are the HF and pair potentials, respectively. We use the approximation that $U$ and $\Delta$ are independent of $\bm{r}$. It is noted that in the superfluid phase (unlike the normal phase) all the HF potentials are equal \cite{Ketterson}. The traditional forms of $\psi(\bm{r}\uparrow)$ and
$\psi(\bm{r}\downarrow)$ are
\begin{equation}
\psi(\bm{r}\uparrow)=\sum_{\bm{k}}(\gamma_{\bm{k}\uparrow}u_{\bm{k}}(\bm{r}\uparrow)-\gamma^{\dagger}_{\bm{k}\downarrow}v_{\bm{k}}^{\star}(\bm{r}\downarrow))
, \ \
\psi(\bm{r}\downarrow)=\sum_{\bm{k}}(\gamma_{\bm{k}\downarrow}u_{\bm{k}}(\bm{r}\downarrow)+\gamma^{\dagger}_{\bm{k}\uparrow}v_{\bm{k}}^{\star}(\bm{r}\downarrow))
\end{equation}
where $\gamma,\gamma^\dagger$ ($u,v$) are the fermionic quasiparticle operators (wavefunctions).
By using these expressions and the commutation relations between $\gamma$,
$\gamma^{\dagger}$ and $H$, one can straightforwardly obtain the
Bogoliubov-de Gennes equations
\begin{eqnarray}
[H(\bm{r}\uparrow)+U(\uparrow)]u(\bm{r}\uparrow)+\Delta
v(\bm{r}\downarrow)=Eu(\bm{r}\uparrow)\nonumber\\
\Delta^{\star}u(\bm{r}\uparrow)-[H(\bm{r}\downarrow)+U(\downarrow)]v(\bm{r}\downarrow)=Ev(\bm{r}\downarrow).
\end{eqnarray}
One obtains the second set of equations by interchanging
$\uparrow$ and $\downarrow$. In the $\alpha$-channel, we take
$u(\bm{r}\uparrow)$ for the particle-like and
$v(\bm{r}\downarrow)$ for the hole-like wavefunctions.

To proceed, let
us take the N-SF interface to be in the $x=0$ plane and introduce the
superscript $s$ ($n$) for the solutions in the SF (N) phase. We also
introduce
$\phi_{\bm{k}(q)}^{\pm}(\bm{r})=\exp[i(\bm{k}_\parallel \cdot
\bm{r}\pm k_{(q)}x)]$ for the N phase, and
$\phi^{\pm(q)}_{\bm{k}}(\bm{r})=\exp[i(\bm{k}_\parallel \cdot
\bm{r}\pm k^{(q)}x)]$ for the SF phase, where $q=p,h$ refers to
particle, hole and $\bm{k}_\parallel$ ($k^{(q)},k_{(q)}$) denotes
the component of the wave vector $\bm{k}$ parallel (perpendicular)
to the interface. Notice that $q$ appears as a subscript
(superscript) for the N (SF) phase throughout our notation. 

From the Bogoliubov-de Gennes equations we obtain the relations,
$k_{(h)}^2=k^{2}_{(p)}+2m_{\uparrow}[U({\uparrow})-m_{r}U({\downarrow})+U_{s}(m_{r}-1)-2\varepsilon]$ and
$k^{(p,h)^2}=k^{2}_{(p)}+2m_{\uparrow}(U({\uparrow})-U_{s}-\xi^{\pm})$,
where $m_{r}=m_{\downarrow}/m_{\uparrow}$ (mass ratio), $2\varepsilon=(E+h_{s})(1+m_{r})+\mu_{s}(1-m_{r})$, and
$\xi^{\pm}=\varepsilon\mp\sqrt{\varepsilon^{2}-m_{r}\Delta^{2}}$. Thus for the N phase we write
\begin{equation}
u^{(n)}_{\bm{k}}(\bm{r}\uparrow)=\sum_{\sigma=\pm}
U^{\sigma}_{\bm{k}(p)}\phi^\sigma_{\bm{k}(p)}(\bm{r}),\ \ \
v^{(n)}_{\bm{k}}(\bm{r}\downarrow)=\sum_{\sigma=\pm}
V^{\sigma}_{\bm{k}(h)}\phi^\sigma_{\bm{k}(h)}(\bm{r}).
\end{equation}
As for the SF phase,
\begin{equation}
u^{(s)}_{\bm{k}}(\bm{r}\uparrow)=\sum_{q,\sigma}U^{\sigma(q)}_{\bm{k}}\phi^{\sigma
(q)}_{\bm{k}}(\bm{r}), \ \ \
v^{(s)}_{\bm{k}}(\bm{r}\downarrow)=\sum_{q,\sigma}V^{\sigma(q)}_{\bm{k}}\phi^{\sigma(q)}_{\bm{k}}(\bm{r})
\end{equation}
where $V^{\sigma(p)}_{\bm{k}}=B U^{\sigma(p)}_{\bm{k}}$ and
$V^{\sigma(h)}_{\bm{k}}=B^{-1} U^{\sigma(h)}_{\bm{k}}$ with
$B=\xi^{+}/\Delta$. The amplitudes $U^{\sigma}_{\bm{k}(p)}$, etc.  are to be
determined by matching the wave functions and their derivatives at
$x=0$, of course \cite{Grefeen}.

Denoting $\xi_{(p)}\equiv k_{(p)}^{2}/2m_{\uparrow}$, for
$\xi^{+}-U({\uparrow})+U_{s}<\xi_{(p)}<\mu_{\uparrow}-U({\uparrow})+E$,
particle-like and hole-like excitations both occur in the SF side,
but Andreev reflection \cite{Andreev1}
 is forbidden. However, for
$\mu_{\uparrow}-U({\uparrow})+E<\xi_{(p)}<2\varepsilon-U({\uparrow})+m_{r}U({\downarrow})-U_{s}(m_{r}-1)$,
we have particle-like and hole-like excitations, as well as normal
and Andreev reflections \cite{Lazaridess1}.
In other regions, the particle
has insufficient energy to excite the SF side and, thus, the transmission
coefficients vanish. We, therefore, restrict our attention to the above two regions, which we shall denote by I and II, respectively. Moreover, our focus is on energies slightly above the transmission threshold
($\varepsilon\approx \sqrt{m_{r}}\Delta$), because we are
considering low temperatures.

Denoting the $x$-component of the current density by
$j_x$, the transmission coefficient is given by
$W=j^{\text{T}}_x/j^{\text{I}}_x$, where the superscripts T and I
refer to the transmitted and incident quasi-particle current
densities, respectively. The general form of $\bm{j}$ (for
$\alpha$-channel) is
\begin{eqnarray}
\bm{j}_\alpha
(\bm{r})=-\frac{i}{2m_{\uparrow}}[u^\star(\bm{r}\uparrow)\nabla
u(\bm{r}\uparrow)-u(\bm{r}\uparrow)\nabla
u^\star(\bm{r}\uparrow)]\nonumber \\
-\frac{i}{2m_{\downarrow}}[-v^\star(\bm{r}\downarrow)\nabla
v(\bm{r}\downarrow)+v(\bm{r}\downarrow)\nabla v^\star(\bm{r}\downarrow)].
\end{eqnarray}
The heat conductivity (for $\alpha$-channel) is given by
\begin{equation}
\kappa= \frac{m_{\uparrow}}{\pi^{2}(1+m_{r})^{2}}\frac{\partial}{\partial T} \int \int
d\xi_{(p)} d\varepsilon\: (\varepsilon-\varepsilon_0)
f(\varepsilon,T)W(\varepsilon,\xi_{(p)})+(\uparrow \rightarrow \downarrow, p\rightarrow h) \label{HC}
\end{equation}
where
$\varepsilon_0=\varepsilon|_{E=0}$ and $f(\varepsilon,T)$ is the Fermi-Dirac distribution, which, in the low temperature limit $T\ll \sqrt {m_r}\Delta$ ($k_B=1$), reduces to
$e^{-T_{\text{m}}/T}$ (up to a proportionality constant), where
\begin{equation}
\frac{3}{2}T_{\text{m}}=2\frac{\sqrt{m_{r}}\Delta-\varepsilon_0}{1+m_{r}}.\label{T} 
\end{equation}
The right hand side is the minimal energy attained by the $\alpha$ spectra, which is positive (in order to have a gapped spectrum) and independent of the HF potentials.

In the mass asymmetric case, analytical calculation of the heat conductivity in the BCS
regime (unlike the deep BCS regime in which the Andreev approximation is valid) is
a formidable task, especially when HF potentials are present. We, therefore, approach the problem numerically and examine
the effect of the HF potentials and mass asymmetry in regions I and II. We
begin by obtaining the allowed range of values for all the relevant parameters in the BCS regime. To this end, we use the following standard relations.  The HF potential of the superfluid phase, obtained by using the fermionic
anticommutation relations for $\gamma$ and $\gamma^{\dagger}$, is given by the number equation
\begin{equation}
U_{s}= Vn_s=\frac{1}{2}V\int\frac{d^{3}k}{(2\pi)^{3}}\:(1-\frac{\zeta_{\bm{k}}}{\sqrt{\zeta_{\bm{k}}^{2}+\Delta^{2}}})
\label{U}
\end{equation}
where
$\zeta_{\bm{k}}=\varepsilon_{\bm{k}}-\mu_{s}=\bm{k}^{2}/2m_{R}-\mu_{s}$. Similarly, the gap equation is given by
\begin{equation}
1=\frac{1}{2}V\int\frac{d^{3}k}{(2\pi)^{3}}\:(\frac{1}{\sqrt{\zeta_{\bm{k}}^{2}+\Delta^{2}}}-\frac{1}{\varepsilon_{\bm{k}}}).\label{gap}
\end{equation}
The above integrals can be calculated using \cite{Papenbrock}
\begin{equation}
\int_{0}^{\infty}
dz\:\frac{z^{\lambda}}{[(z-1)^{2}+x^{2}]^{1/2}}=-\frac{\pi}{\sin \pi\lambda}\:(1+x^{2})^{\lambda/2}P_{\lambda}(\frac{-1}{\sqrt{1+x^{2}}})
\end{equation}
where $P_{\lambda}$ is the Legendre function. Equations (\ref{U}) and (\ref{gap}), thus, yield
\begin{equation}
\frac{1}{m_{R}a^{2}}=-\frac{2\mu_{s}}{\varsigma}P_{1/2}^2(\varsigma),\ \ \
U_{s}=-\mu_{s}\left[1-\frac{P_{3/2}(\varsigma)}{\varsigma P_{1/2}(\varsigma)}\right] \label{range}
\end{equation}
where $\varsigma=-[1+(\Delta/\mu_{s})^{2}]^{-1/2}$. Since $n_{s}=k_{\text{F}}^{3}/3\pi^{2}$, using (\ref{range})
we find
\begin{equation}
(k_{\text{F}}a)^{-3}=\frac{4}{3\pi} \frac{P_{1/2}^3(\varsigma)}{P_{3/2}(\varsigma)-\varsigma P_{1/2}(\varsigma)}.
\end{equation}
This relationship determines the allowed values of $\varsigma$ by fixing $1/k_{\text{F}}a$ in the BCS regime. Through (\ref{range}) we thus find $\mu_s$, $U_s$, $\Delta$, and the latter yields $h_s$ via the Clogston limit. We, therefore, have $\mu_{\uparrow}$ and
$\mu_{\downarrow}$ as well. In the normal phase we similarly find \cite{Lazaridess1}
\begin{equation}
U(\uparrow)=-\frac{4\sqrt{2}m_{R}a^{5}}{3\pi}[m_{\downarrow}(\mu_{\downarrow}-U(\downarrow))]^{3/2},\ \ \ U(\downarrow)=-\frac{4\sqrt{2}m_{R}a^{5}}{3\pi}[m_{\uparrow}(\mu_{\uparrow}-U(\uparrow))]^{3/2} \label{Us}
\end{equation}
which yield the allowed values of $U(\uparrow)$ and $U(\downarrow)$ in the BCS regime. For $m_r$ less than a cut-off value $M$ (which depends on the interaction strength), we find two solutions for each $U(i)$, satisfying $U_{s}<U(i)<0 $ (for $m_r>M$, no real solution exists).  Since the interactions are attractive the upper bound on the potentials is obvious. The lower bound implies that the density of the SF phase (the core region) exceeds that of the N phase. This reconciles with the fact that a N-SF interface exists, separating an unpolarized SF from a partially polarized N phase. Therefore, we take $1\leq m_r<M $ for the allowed range of values of $m_r$ in the BCS regime.

\section{Results and disscusion}
\begin{figure}
\includegraphics{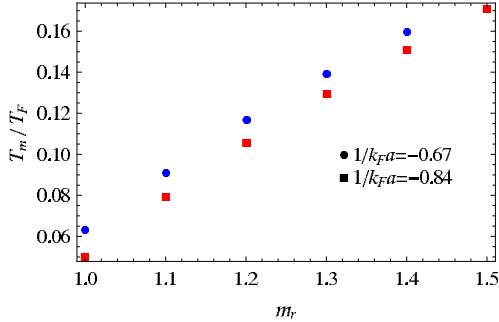}
\caption{(Color online) $T_{\text{m}}$ versus $m_{r}$.}\label{fig1}
\end{figure}

The relevant physical parameters in a dilute Fermi gas are the temperature, the mass ratio, and the interaction strength. It is, therefore, important to find how physical quantities depend on these parameters \cite{Zhang}.  For the heat conductivity, (\ref{HC}) is found to give
\begin{equation}
\frac{\kappa}{\kappa_{\text{N}}}= G\left(\frac{T_{\text{F}}}{T}\right)^{3/2} e^{-\frac{3}{2}\frac{T_{\text{m}}}{T}}\label{formula}
\end{equation}
where $T_{\text{m}}$ is defined by (\ref{T}), $\kappa_{\text{N}}=T(\mu_\uparrow m_\uparrow+\mu_\downarrow m_\downarrow)/\pi^2$ is the heat conductivity of the N phase, and $T_{\text{F}}$ is the Fermi temperature. The functional forms of $G(\frac{1}{k_{\text{F}}a},m_r)$ and $T_{\text{m}}(\frac{1}{k_{\text{F}}a},m_r)$ will be discussed shortly. We note that, as a consequence of (\ref{formula}), $\kappa/\kappa_{\text{N}}$ starts to drop from its maximum value at $T=T_{\text{m}}$ with decreasing temperature. This signifies a build up of temperature difference across the interface, which be understood as follows. According to (\ref{T}), at $T=T_{\text{m}}$, the mean kinetic energy of the particles is just sufficient to overcome the SF gap. We, therefore, expect a blockage of energy transfer at lower temperatures, resulting in the reflection of particles from the interface and hence a drop in the interface conductivity. $T_{\text{m}}$ is an increasing function of $m_r$, which, for fixed $m_r$, increases with the interaction strength too (Fig. \ref{fig1}). 

\begin{figure}
\includegraphics{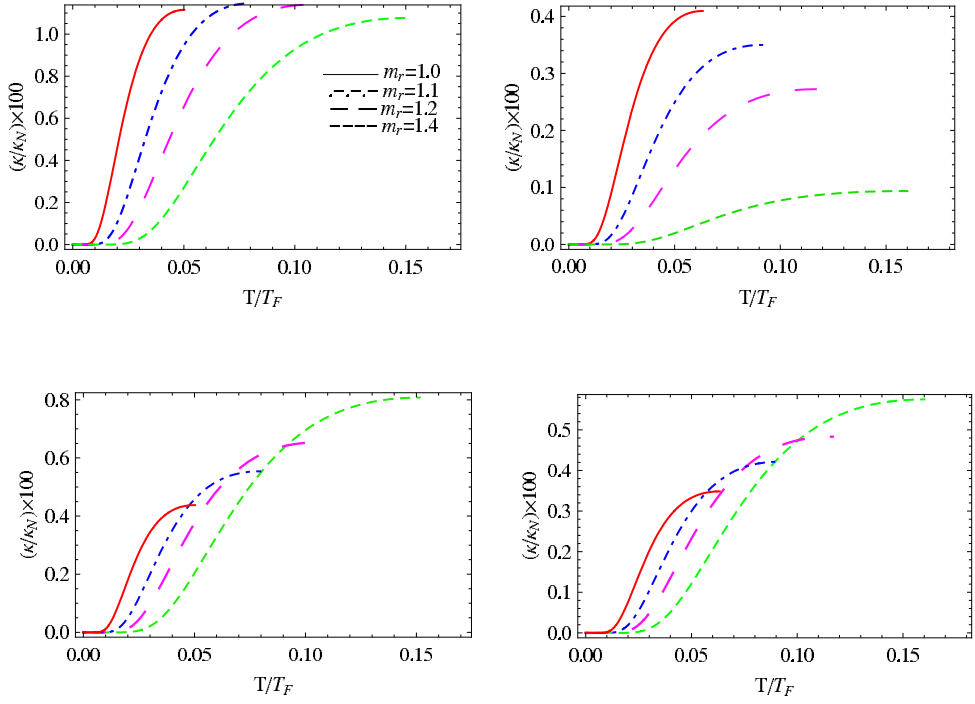}
\caption{Interface conductivity versus temperature for $1/k_{\text{F}}a=-0.84$ (left) and -0.67 (right), with (top) and without (bottom) HF potentials.} \label{fig2}
\end{figure}

Fig. \ref{fig2} shows typical results for the temperature variation of $\kappa/\kappa_{\text{N}}$ using $1/k_{\text{F}}a=-0.84, -0.67$.
As seen, for fixed $m_{r}$, the larger the
absolute value of $1/k_{\text{F}}a$ (i.e., the weaker the interaction), the larger is the heat conductivity. 
Also, in the presence of HF potentials, the maximum value of $\kappa$ is almost the same for all $m_r$ for sufficiently weak interactions.   

\begin{figure}
\includegraphics{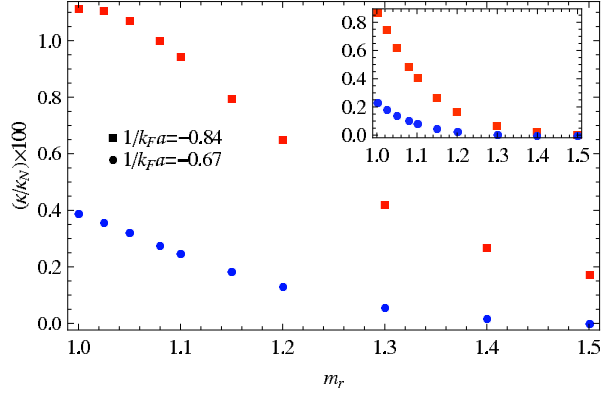}
\caption{(Color online) Interface conductivity versus mass ratio at $T/T_{\text{F}}=0.05$ and 0.03 (inset).}\label{fig3}
\end{figure}

Furthermore, the $\kappa/\kappa_{\text{N}}$ at fixed temperature decreases
with $m_{r}$ (Fig. \ref{fig3}), resulting in an increase in the temperature difference across the interface. 
This means that the characteristic relaxation time increases with $m_r$. Note that for sufficiently high values of the mass ratio, $\kappa/\kappa_{\text{N}}$ is independent of the interaction strength, provided $T/T_{\text{F}}\lesssim  0.03$.

\begin{figure}
\includegraphics{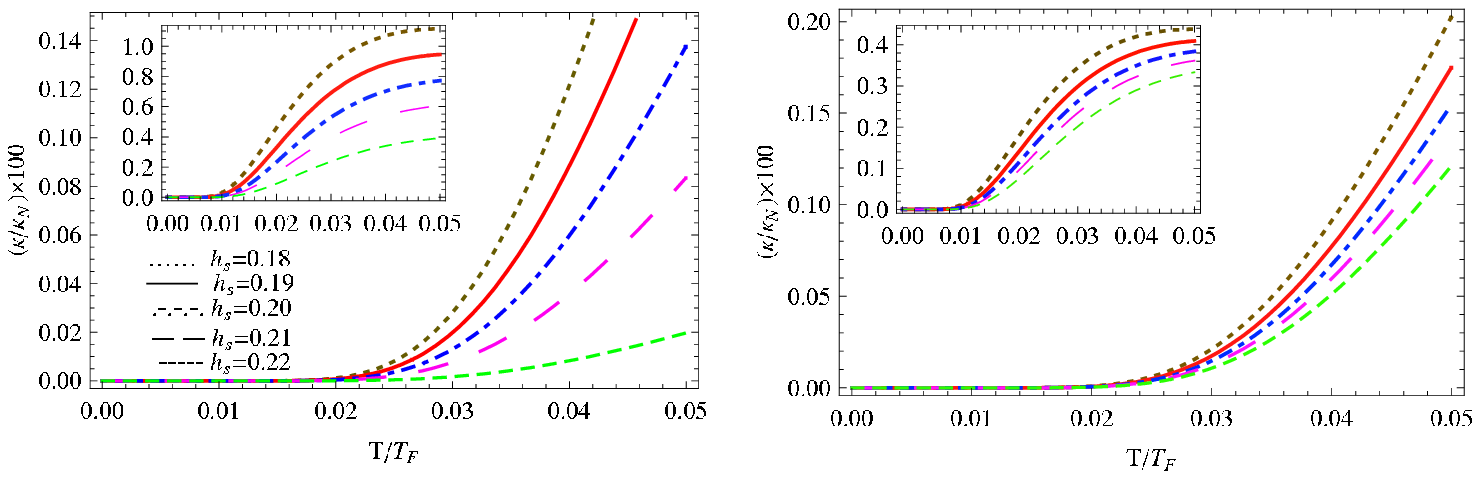}
\caption{(Color online) Interface conductivity versus temperature for $m_{r}=1$ (inset) and 1.4, with (left) and without (right) HF potentials.}\label{fig4}
\end{figure}

Fig. \ref{fig4} shows typical curves of 
$\kappa/\kappa_{\text{N}}$ versus $T$ at fixed $h_{s}$, for $m_{r}=1, 1.4$.
At fixed $T$, $\kappa/\kappa_{\text{N}}$ decreases with increasing $h_{s}$, because of the growing interaction strength. The effect of $h_s$ (which can be controlled by the species population
imbalance) on $\kappa$ is more pronounced in the presence of HF potentials, because the latter affect the threshold line ($\varepsilon= \Delta\sqrt{m_{r}}$) by changing $\varepsilon$. However, as seen, the role of $h_s$ diminishes at sufficiently low temperatures.
 
\begin{figure}
\includegraphics{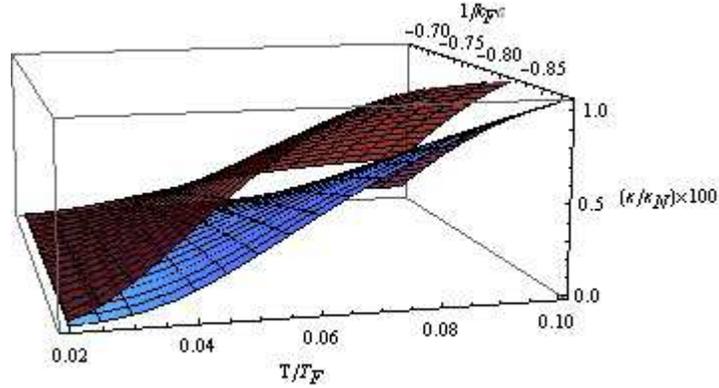}
\caption{(Color online) Dependence of interface conductivity on temperature and interaction strength for $m_{r}=1.2$ (top) and 1.4 (bottom).}\label{fig5}
\end{figure}

The functional forms of $G$ and $T_{\text{m}}$ in (\ref{formula}) have been determined by the method of least-squares fit. They are valid for the whole BCS regime and reproduce the above results very accurately:
\begin{eqnarray}
G(\frac{1}{k_{\text{F}}a},m_r)=g_0(m_r)+ \frac{1}{k_{\text{F}}a}g_1(m_r)+ (\frac{1}{k_{\text{F}}a})^2g_2(m_r)\nonumber \\
\frac{3}{2}T_{\text{m}}(\frac{1}{k_{\text{F}}a},m_r)=\frac{T_{\text{F}}m_r}{1+m_r}\left[t_0(m_r)+\frac{1}{k_{\text{F}}a} t_1(m_r)+ (\frac{1}{k_{\text{F}}a})^2t_2(m_r)\right]
\end{eqnarray}
where
\begin{eqnarray}
g_0=-0.51+\frac{0.93}{\surd{m_r}}-\frac{0.43}{m_r},\ 
g_1=-1.00+\frac{1.78}{\surd{m_r}}-\frac{0.79}{m_r},\ 
g_2=-0.44+\frac{0.75}{\surd{m_r}}-\frac{0.32}{m_r}\nonumber \\
t_0=0.16+\frac{1.38}{\surd{m_r}}-\frac{1.14}{m_r},\ 
t_1=-0.99+\frac{1.35}{\surd{m_r}}+\frac{0.03}{m_r},\ 
t_2=-0.34+\frac{0.39}{\surd{m_r}}+\frac{0.07}{m_r}. \nonumber
\end{eqnarray}
The resulting functional dependence of $\kappa/\kappa_{\text{N}}$ on the temperature and interaction strength is depicted in Fig. \ref{fig5}.  

\begin{figure}
\includegraphics{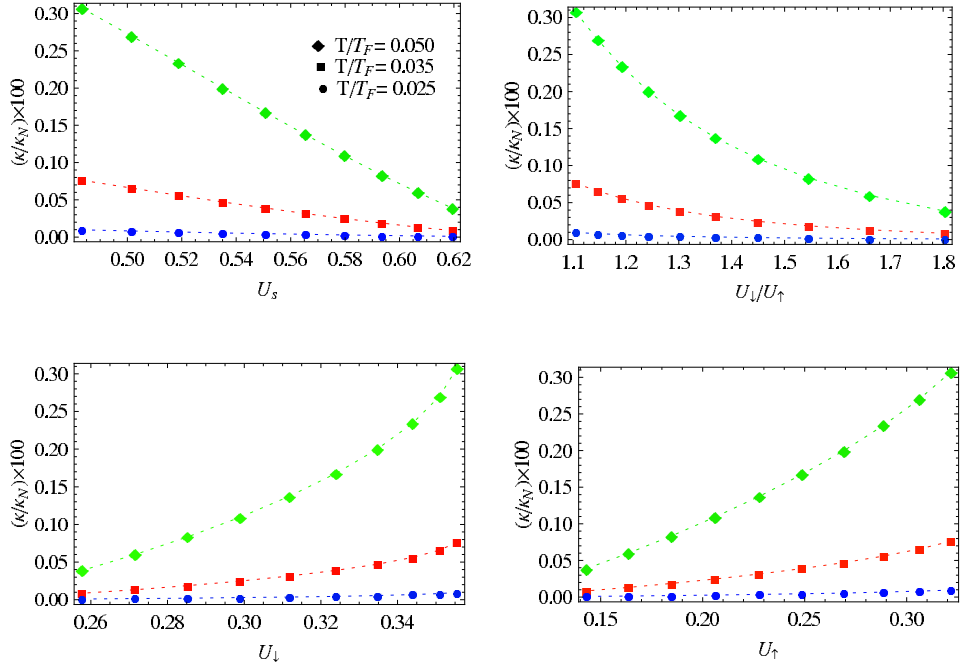}
\caption{(Color online) Interface conductivity versus HF potentials for $m_{r}=1.4$.} \label{fig6}
\end{figure}

The curves of $\kappa/\kappa_{\text{N}}$ versus HF potentials at constant temperature are shown in Fig. \ref{fig6}.
As seen, the role of the potentials becomes more important as the temperature increases. Also, the heat
conductivity decreases with $U_{\downarrow}/U_{\uparrow}$, which is understood because of the resulting increase in the scattering length (and, hence, cross section). 

As a subsidiary result, we may also point out that our numerical calculations show that the role of the incident particles (from the N side) with energies in region I (where Andreev reflection does not occur) 
is much more important in $\kappa$ than that of the incident particles/holes
with energies in region II.

\subsection*{The mixture $^{6}\text{Li}$-$^{40}\text{K}$}

Here, we consider in more detail the particular case of the $^6\text{Li}$-$^{40}\text{K}$
mixture ($m_r=6.7$), due to its importance in experimental and
theoretical studies. Since $T_{\text{m}}$ increases with $m_r$ (Fig. \ref{fig1}), for larger values of the mass ratio such as here, the range of relevant temperatures increases. Hence, it would be more appropriate to take the temperature dependence of $\Delta$ into account. We have \cite{Abrikosov}
\begin{equation}
\frac{\Delta(T)}{\Delta}-1\propto(8-\frac{T}{\Delta})\sqrt{\frac{T}{\Delta}} e^{-\Delta/T}\label{delta}
\end{equation}
where $\Delta$ is the zero temperature limit considered in our previous calculations. Also, we take the distribution function $f(\varepsilon,T)$ in (\ref{HC}) to be the exact Fermi-Dirac distribution instead of its approximate low-temperature form. (However, for simplicity, we take HF potentials to be zero.)

\begin{figure}
\includegraphics{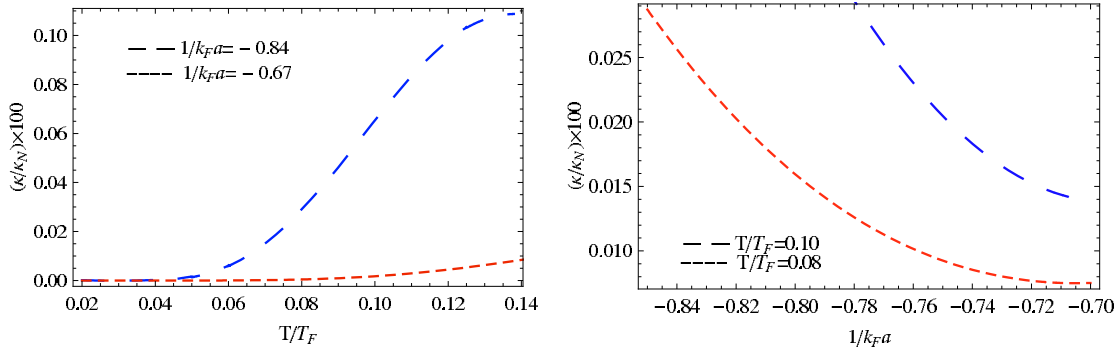}
\caption{(Color online) Interface conductivity versus temperature and interaction strength for $^{6}\text{Li}$-$^{40}\text{K}$ mixture.}  \label{fig7}
\end{figure}

\begin{figure}
\includegraphics{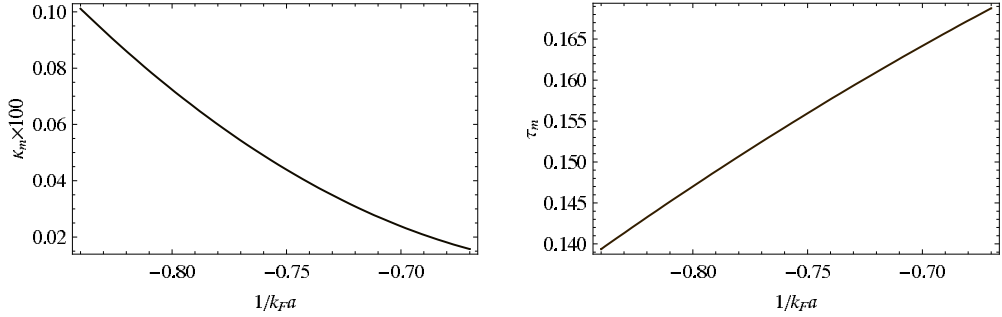}
\caption{$\kappa_{\text{m}}$ and $\tau_{\text{m}}$ versus interaction strength for $^{6}\text{Li}$-$^{40}\text{K}$ mixture.}
 \label{fig8}
\end{figure}

We find the following analytic expression for the transmission coefficient of region I:
\begin{eqnarray}
W_{I}=8(\varepsilon-\varepsilon_0)[2\sqrt{m_{r}}\Delta(T)(\varepsilon-\sqrt{m_r}\Delta(T))]^{\frac{1}{2}}\times \ \ \ \ \ \ \ \ \ \ \ \ \ \ \ \ \ \ \ \nonumber\\
\frac{\sqrt{\chi(\chi-1)}[(1+m_{r}^{2})(2\chi-3)+2(m^{2}_{r}-1)\sqrt{(\chi-1)(\chi-2)}]}{\left[(m_{r}-1)+(\sqrt{\chi-2}-\sqrt{\chi})[(m_{r}-1)\sqrt{\chi}-(m_{r}+1)\sqrt{\chi-1}]\right]^{2}}
\end{eqnarray}
where $\chi=\xi_{(p)}/\Delta(T)\sqrt{m_{r}}$. By using equations (\ref{delta}), (\ref{range}), and the definition of $\varsigma$, we can obtain the temperature dependence of $\mu_{s}$. The interface conductivity (\ref{HC}) is then obtained via numerical interpolation (Fig. \ref{fig7}).
Similarly, the functional forms of $\kappa_{\text{m}}$ and $\tau_{\text{m}}$ (the maximum values of $\kappa/\kappa_{\text{N}}$ and $T/T_{\text{F}}$, respectively) have been determined by the method of least-squares fit. They are
\begin{eqnarray}
\kappa_{\text{m}}(\frac{1}{k_{\text{F}}a})=0.006+0.02 \frac{1}{k_{\text{F}}a}+0.02 (\frac{1}{k_{\text{F}}a})^2\nonumber \\
\tau_{\text{m}}(\frac{1}{k_{\text{F}}a})=0.2-0.04\frac{1}{k_{\text{F}}a}-0.14 (\frac{1}{k_{\text{F}}a})^2.
\end{eqnarray}
Fig. \ref{fig8} shows the graphs of $\kappa_{\text{m}}$ and $\tau_{\text{m}}$ versus the interaction strength. 

It is noteworthy that, since $h_s<\sqrt{m_{r}} \Delta(T)<\mu_s$, where the second inequality sign is owing to the fact that $\kappa$ is real, the condition for Clogston limit ($h_s \ll \mu_s$) is satisfied more stringently as $m_r$ increases.

\end{document}